\documentclass[11pt,a4paper]{article}
\usepackage{graphicx}
\usepackage{amsmath}
\usepackage{amsfonts}

\title{A quadrature-based eigensolver with a Krylov subspace method for shifted linear systems for Hermitian eigenproblems in lattice QCD}
\author{H.~Ohno$^{1,2}$, Y.~Kuramashi$^{1,3}$, T.~Sakurai$^{2}$ and H.~Tadano$^{2,3}$}
\date{\empty}

\begin{document}
\maketitle

\vspace{-2em}
\begin{center}
\textit{$^1$Graduate School of Pure and Applied Sciences, University of Tsukuba, Tsukuba, Ibaraki 305-8571, Japan \\
$^2$Graduate School of Systems and Information Engineering, University of Tsukuba, Tsukuba, Ibaraki 305-8573, Japan \\
$^3$Center for Computational Sciences, University of Tsukuba, Tsukuba, Ibaraki 305-8577, Japan}
\end{center}

\begin{abstract}
We consider a quadrature-based eigensolver to find eigenpairs of Hermitian matrices arising in lattice quantum chromodynamics.
To reduce the computational cost for finding eigenpairs of such Hermitian matrices, we propose a new technique for solving shifted
linear systems with complex shifts by means of the shifted CG method. Furthermore using integration paths along horizontal lines corresponding
to the real axis of the complex plane, the number of iterations for the shifted CG method is also reduced.
Some numerical experiments illustrate the accuracy and efficiency of the proposed method by comparison with a conventional method.
\end{abstract}

\section{Introduction}
Eigenproblems arise in many scientific applications and in some cases, only a limited set of eigenpairs is needed.
For example, to calculate all-to-all propagators in lattice quantum chromodynamics (QCD) \cite{all-to-all}, it is known that the contribution
of some low-lying eigenvalues of a large sparse Hermitian matrix called ``Hermitian fermion matrix" is dominant. 

For such eigenproblems, the Implicitly Restarted Lanczos method or generally, the Implicitly Restarted Arnoldi method (IRAM) \cite{IRAM}
is one of the conventional choices. On the other hand, to find eigenvalues in a given region and corresponding eigenvectors with contour
integrations, the Sakurai-Sugiura (SS) method \cite{SS} has been proposed. The SS method translates a problem of finding eigenvalues
in a domain surrounded by an integration path into a problem of solving systems of linear equations for some matrices with shifts
corresponding to quadrature points of the contour integration. Therefore solving shifted linear systems efficiently plays an important role
of high performance of the SS method. In this paper, we improve the SS method by reducing computational cost for solving shifted linear systems.

To solve such shifted linear systems efficiently, there are some ways. One is solving each linear system in parallel
since all shifted linear systems arising in the SS method can be solved independently. This high parallelism is one of the important feature
of the SS method. Another way is reducing matrix-vector multiplications in a Krylov subspace linear solver.
We consider the latter in this paper. 

To that end, we propose following two ideas: first we adopt a Krylov subspace method for shifted linear systems \cite{shifted_Krylov}
which needs matrix-vector multiplications only for one seed linear system to solve all shifted linear systems because of shift invariance of
Krylov subspace. Moreover we show that the shifted CG method, which needs no more than one matrix-vector multiplication in each iteration, can be
applied to solve such shifted linear systems if a coefficient matrix of a seed linear system is Hermite although coefficient matrices of
the other shifted systems are non-Hermite. We show the detail in Section 2.
Second, in Section 3, we consider appropriate configurations of quadrature points for less iterations for a Krylov subspace solver
because the number of iterations depends on a shift parameter corresponding to each quadrature point. Section 4 shows some numerical test
to investigate the properties of the proposed method and its efficiency by comparison with PARPACK \cite{PARPACK}, the software package
to solve eigenvalue problems with IRAM in parallel, with a Hermitian fermion matrix of lattice QCD and Section 5 concludes.

\section{The SS method with a Krylov subspace method for shifted linear systems}
\subsection{The SS method}
We consider an eigenvalue problem $A \textrm{\boldmath $x$} = \lambda \textrm{\boldmath $x$}$ where $A \in \mathbb{C}^{n\times n}$
is a Hermitian matrix and $(\lambda,\textrm{\boldmath $x$})$ is an eigenpair of $A$. Let $\Gamma$ be a positively oriented closed Jordan curve
in the complex plane and introduce a contour integration
\begin{equation}\label{s_k}
\textrm{\boldmath $s$}_k \equiv \frac{1}{2\pi i}\int_{\Gamma}z^k(zI-A)^{-1}\textrm{\boldmath $v$}dz,\quad k=0,1,\cdots,
\end{equation}
where $I$ is an $n\times n$ unit matrix and $\textrm{\boldmath $v$} \in \mathbb{C}^{n}$ is any nonzero vector. According to the residue theorem,
$\textrm{\boldmath $s$}_k$ has only the contribution corresponding to eigenvalues inside $\Gamma$. 

In the moment based method \cite{SS}, a moment $\mu_k \equiv \textrm{\boldmath $v$}^{\mathrm{H}}\textrm{\boldmath $s$}_k$
is defined and let the Hankel matrix $H_m \in \mathbb{C}^{m\times m}$ and the shifted Hankel matrix $H^{<}_m \in \mathbb{C}^{m\times m}$ be
\begin{equation*}
H_m \equiv [\mu_{i+j-2}]^{m}_{i,j=1},\quad H^{<}_m \equiv [\mu_{i+j-1}]^{m}_{i,j=1},
\end{equation*}
respectively, where $m$ is the number of eigenvalues inside $\Gamma$. Here let
$S \equiv [\textrm{\boldmath $s$}_0,\cdots,\textrm{\boldmath $s$}_{m-1}] \in \mathbb{C}^{n\times m}$, eigenvalues of the pencil
$(H^{<}_m,H_m)$ are given by $\lambda_1,\cdots, \lambda_m$ and an eigenvector corresponding to $\lambda_l$ is given by
$\textrm{\boldmath $x$}_l = S\textrm{\boldmath $u$}_l$ where $\textrm{\boldmath $u$}_l$ is an eigenvector of ($H^{<}_m,H_m$).

On the other hand, in a Rayleigh-Ritz type approach \cite{CIRR}, constructing an orthonormal basis $Q \in \mathbb{C}^{n\times m}$
via the orthogonalization of $S$, approximate eigenvalues are given by the Ritz values of a projected matrix pencil $(\tilde{A},\tilde{B})$
where $\tilde{A} \equiv Q^{\mathrm{H}}AQ \in \mathbb{C}^{m\times m}$ and $\tilde{B} \equiv Q^{\mathrm{H}}Q \in \mathbb{C}^{m\times m}$, respectively,
and corresponding eigenvectors are given by $\textrm{\boldmath $x$}_l = Q\textrm{\boldmath $w$}_l$ where $\textrm{\boldmath $w$}_l$
is an eigenvector of $(\tilde{A},\tilde{B})$.

The Rayleigh-Ritz projection method is rather accurate than the moment-based method, however there is trade-off between accuracy and
memory consumption.

To calculate (\ref{s_k}) numerically, the $N$-point trapezoidal rule is applied and we approximate $\textrm{\boldmath $s$}_k$ by
\begin{equation}\label{s_hat_k}
\hat{\textrm{\boldmath $s$}}_k = \sum^{N-1}_{j=0} w_j \zeta^k_j(z_{j}I-A)^{-1}\textrm{\boldmath $v$},
\end{equation}
where $z_j$ and $w_j$ are a quadrature point and a weight, respectively, and $\zeta_j\equiv (z_j-\gamma)/\rho$ is a normalized quadrature
point satisfying the condition $-1 \leq \mathrm{Re}\;\zeta_j \leq 1$ with a shift parameter $\gamma \in \mathbb{C}$ and a scale parameter $\rho > 0$.
In the case of an integration on a circle $C$ with a center $\gamma$ and a radius $\rho$, a quadrature point and a weight are defined by
\begin{equation*}
z_j=\gamma + \rho\; e^{\frac{2 \pi i}{N}(j+1/2)},\quad j=0,1,\cdots,N-1,
\end{equation*}
and 
\begin{equation*}
w_j=\frac{z_j-\gamma}{N},\quad j=0,1,\cdots,N-1,
\end{equation*}
respectively. Here let $\eta_l \equiv (\lambda_l-\gamma)/\rho$ and suppose $\textrm{\boldmath $v$} = \sum_{l}\alpha_{l}\textrm{\boldmath $x$}_l$,
we can rewrite (\ref{s_hat_k}) as
$\hat{\textrm{\boldmath $s$}}_k = \sum_{l}f_k(\eta_{l}) \alpha_{l}\textrm{\boldmath $x$}_l/\rho$ by means of a filter function defined by
\begin{equation}\label{filter_func}
f_k(x) \equiv \sum^{N-1}_{j=0} \frac{w_j\zeta^{k}_{j}}{\zeta_{j}-x}.
\end{equation}
In the case of a unit circle, it has been shown that $f_k(x)=x^{k}/(1+x^{N})$ \cite{filter} and it suppresses as $O(|x|^{-N+k})$ outside the circle.
This means that $\hat{\textrm{\boldmath $s$}}_k$ has nonnegligible contribution corresponding to the eigenvalues outside the circle due to the approximation
of the contour integration. When we construct $\hat{S} \equiv [\hat{\textrm{\boldmath $s$}}_0,\cdots,\hat{\textrm{\boldmath $s$}}_{M-1}]$, $M$ should be more than $m$.

Note that a block version of the SS method is proposed in \cite{filter}, i.e. $S$ is extended to $\mathbb{C}^{n \times (M \times L)}$ with
$L$ different arbitrary nonzero vectors $\textrm{\boldmath $v$}_1,\cdots,\textrm{\boldmath $v$}_L$. Using this method, we can obtain $L$ degenerate
eigenvalues and it is known that the accuracy is higher than just increasing $M$.

\subsection{The SS method with the shifted CG method}
To calculate (\ref{s_hat_k}), shifted linear systems such that
\begin{equation}\label{linear_eq}
(z_j I-A)\textrm{\boldmath $y$}_j = \textrm{\boldmath $v$}
\end{equation}
should be solved for each quadrature point $z_j$. 
In what follows, we propose some ideas to reduce the computational cost to solving (\ref{linear_eq}) with the shifted CG method \cite{shifted_Krylov}. 

It is known that there is a shift invariance of Krylov subspace
$\mathcal{K}_k(A,\textrm{\boldmath $r$}_0)\equiv \mathrm{span}(\textrm{\boldmath $r$}_0, A\textrm{\boldmath $r$}_0,\cdots, A^{k-1}\textrm{\boldmath $r$}_0)$
with any shift $\sigma \in \mathbb{C}$ such that
\begin{equation}\label{shift}
\mathcal{K}_k(A,\textrm{\boldmath $r$}_0) = \mathcal{K}_k(A + \sigma I,\textrm{\boldmath $r$}_0).
\end{equation}
In a Krylov subspace linear solver such as the CG method, matrix-vector multiplications should be performed to update a residual vector
$\textrm{\boldmath $r$}_k \in \mathcal{K}_{k+1}(A,\textrm{\boldmath $r$}_0)$. Because of (\ref{shift}), a residual vector
$\textrm{\boldmath $r$}^{\sigma}_k \in \mathcal{K}_{k+1}(A+\sigma I,\textrm{\boldmath $r$}_0)$ corresponding to a matrix
$A+\sigma I$ can be given by
\begin{equation*}
\textrm{\boldmath $r$}^{\sigma}_k = \xi^{\sigma}_k \textrm{\boldmath $r$}_k,
\end{equation*}
with some scalar $\xi^{\sigma}_k$, namely once a residual vector of a seed linear system $\textrm{\boldmath $r$}_k$ is calculated
with matrix-vector multiplications, residual vectors of any other shifted linear systems $\textrm{\boldmath $r$}^{\sigma}_k$,
what is more, corresponding solution vectors are given
without additional matrix-vector multiplications. Suppose the computational cost of matrix-vector multiplications is dominant,
the computational cost of the SS method is drastically reduced by $1/N$.

In addition, we show that only one matrix-vector multiplication in each iteration is needed to solve shifted linear systems when a coefficient
matrix consists of a Hermitian matrix with any shift.
Consider the BiCG method to solve a system of linear equations with a coefficient matrix $\sigma I - A$.
The BiCG method needs two matrix-vector multiplications in each iteration to update a residual vector
$\textrm{\boldmath $r$}_k \in \mathcal{K}_{k+1}(\sigma I -A,\textrm{\boldmath $r$}_0)$ and its shadow residual vector
$\textrm{\boldmath $r$}^{*}_k \in \mathcal{K}_{k+1}((\sigma I - A)^{\mathrm{H}},\textrm{\boldmath $r$}^{*}_0)$.
When $A$ is Hermite, we find that 
\begin{equation*}
(\sigma I - A)^{\mathrm{H}}  = (\sigma I - A) + (\bar{\sigma}-\sigma)I.
\end{equation*}
Because of the shift invariance (\ref{shift}), $\textrm{\boldmath $r$}^{*}_k$ is calculated by $\textrm{\boldmath $r$}_k$ without matrix-vector multiplications
and accordingly, the computational cost is reduced by half, if $\textrm{\boldmath $r$}^{*}_0 = \textrm{\boldmath $r$}_0 $.
Applying this technique to the shifted BiCG method, we can solve many shifted linear systems with only one matrix-vector multiplication in each iteration.
If a shift $\sigma$ for the seed system is real, $(\sigma I - A)^{\mathrm{H}} = \sigma I - A$, i.e. the seed system and its shadow system are coincident.
In this case, we can apply the CG method to solve the seed system.
This means that any shifted linear systems with arbitrary complex shift can be solved with the shifted CG method when a coefficient matrix corresponding to
a seed system consists of a Hermitian matrix with some real shift.
For the above reason, we adopt the shifted CG method in this paper.

\section{Integrations along straight lines}
\begin{figure}[tbp]
\begin{center}
\includegraphics[width=6cm,angle=-90]{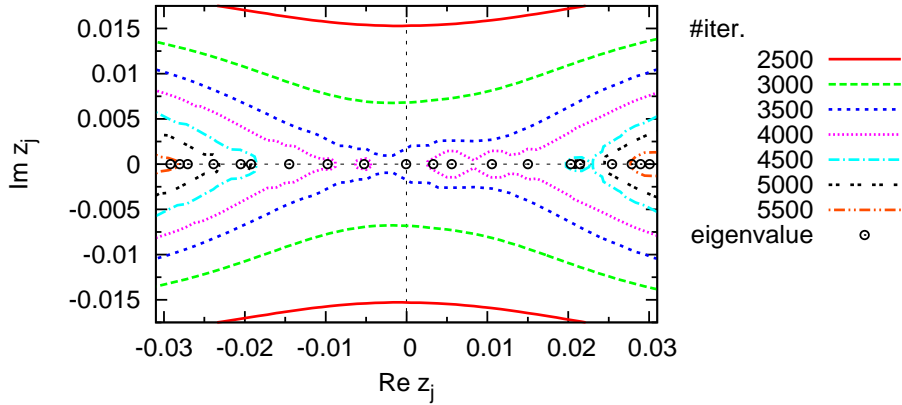}
\caption{The distribution of eigenvalues for the Hermitian fermion matrix which we use in Section 4
and the contour of the number of iterations for the shifted CG method for $z_j I -A$ with tolerance
for the relative residual $||\textrm{\boldmath $r$}_k||_2/||\textrm{\boldmath $v$}||_2 \leq 10^{-12}$.}
\label{iter}
\end{center}
\end{figure}

Empirically, the number of iterations for the shifted CG method depends on distribution of eigenvalues near zero. In terms of shifted matrix
$z_j I - A$, the number of iterations for the shifted CG method tends to increase if the number of eigenvalues of $A$ close to $z_j$
becomes larger. Fig. \ref{iter} shows the distribution of eigenvalues for the Hermitian fermion matrix which we use in Section 4 and the contour of
the number of iterations for the shifted CG method to solve a shifted matrix $z_j I - A$ with tolerance for the relative residual
$||\textrm{\boldmath $r$}_k||_2/||\textrm{\boldmath $v$}||_2 \leq 10^{-12}$. Actually, the number of iterations for the shifted CG method gets
larger as $z_j$ becomes closer to the real axis and its absolute value of real part becomes larger, which is consistent with the expectation from
the distribution of eigenvalues as mentioned above. Then to reduce the number of iterations for the shifted CG method, quadrature points should be
as far from the real axis as possible as accuracy of eigenpairs is enough for applications.

\begin{figure}[tbp]
\begin{center}
\includegraphics[width=6cm,angle=-90]{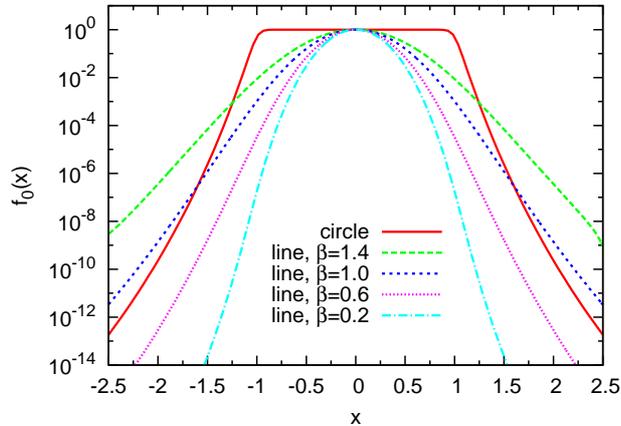}
\caption{Filter functions $f_0(x)$ for $N=32$ quadrature points on $\mathbb{L}^{\pm}$ with $\beta=0.2,0.6,1.0,1.4$.
For comparison, $f_0(x)$ in the case of $C$ is also shown as a solid line.}
\label{filter}
\end{center}
\end{figure}

In order to control the number of iterations for the shifted CG method and accuracy of eigenpairs, we introduce a new integration path as follows:
let $\mathbb{L}^{\pm}$ be two horizontal lines such that
\begin{equation*}
\mathbb{L}^{\pm}:z=\gamma + \rho(x\pm i\beta),\; -1 \leq x \leq 1,
\end{equation*}
where $\gamma$ is real.
Then $N/2$ equally-spaced quadrature points $z_0,z_2,\cdots,z_{N-2}$ and $z_1,z_3,\cdots,z_{N-1}$ are located on
$\mathbb{L}^{+}$ and $\mathbb{L}^{-}$, respectively. Using this integration path, distance between a quadrature point and the real axis depends
only on $\beta\rho$ unlike the case of $C$. In this case, a set of weights for a integration $\{w_0,w_1,\cdots,w_{N-1}\}$
is defined by solving following linear equations,
\begin{equation*}
\sum^{N-1}_{j=0}w_j\zeta^{k-1}_j = \left \{
\begin{array}{ll}
1, & k=0\\
0, & k=1,\cdots, N-1 \\
\end{array}
\right. .
\end{equation*}

Fig. \ref{filter} shows filter functions $f_0(x)$ defined by (\ref{filter_func}) for $N=32$ quadrature points on $\mathbb{L}^{\pm}$ with
$\beta=0.2,0.6,1.0,1.4$. For comparison, $f_0(x)$ in the case of $C$ is also shown. In the case of $C$, $f_0(x)$ has
a plateau in [-1,1] and exponentially suppresses in the other region. However in the case of $\mathbb{L}^{\pm}$, there is no plateau and the slope
of dumping parts depends on $\beta$ which controls the size of the gap between $\mathbb{L}^{+}$ and $\mathbb{L}^{-}$. Therefore it is expected that
the accuracy of eigenvalues is the highest right in the middle of $\mathbb{L}^{\pm}$ and gets lower at the edges, so we accept the eigenvalues only in the
rectangular formed by $N^{\prime}$ quadrature points near the center $\gamma$ where $N^{\prime} \geq 4$ is an even number.

Note that when we introduce more than two integration paths lying next to each other $\mathbb{L}^{\pm}_1,\mathbb{L}^{\pm}_2,\cdots$
which have the same $N$, $\rho$ and $\beta$, quadrature points can be reused by just shifting them from one integration path
$\mathbb{L}^{\pm}_k$ to another $\mathbb{L}^{\pm}_{k+1}$ by $N^{\prime}-2$ points. This is advantage of using the integration path
$\mathbb{L}^{\pm}$.

\section{Numerical experiments}
A Hermitian fermion matrix is defined as $A = \gamma_5(I-\kappa D)$ where 
$\kappa$ is a hopping parameter
and $D$ is a complex non-symmetric sparse matrix
explained in \cite{fermion}.
$\gamma_5$ is one of the Dirac $\gamma$ matrices.
We employ the lattice size $12^3\times 24$ which corresponds to a $497,664$ 
dimensional matrix with $25,380,864$ nonzero components.
Our choice of $\kappa=0.13600$ is rather close to the critical value 
$\kappa_c=0.136116(8)$. 

Our experiments are carried out on a single node of T2K-TSUKUBA which has totally 648 nodes providing 95.4 Tflops of computing capability.
Each node has 4 sets of a 2.3 GHz Quad-Core AMD Opteron Model 8356 processor and a 8 GBytes DDR2-667 memory.

Let us first compare the efficiency of the SS method 
between the conventional integration path $C$ and the new integration path
$\mathbb{L}^{\pm}$
by calculating low-lying 6 eigenvalues and corresponding eigenvectors.
In both cases, we set $\gamma=0.0004$, $\rho=0.0102$, $N=32$ and $M=24$.
The gap size between $\mathbb{L}^{+}$ and $\mathbb{L}^{-}$ is varied 
as $\beta = 0.2,0.6,1.0$.
We employ the shifted CG method with the stopping criterion 
for the relative residual 
$||\textrm{\boldmath $r$}_k||_2/||\textrm{\boldmath $v$}||_2 \leq 10^{-12}$
choosing a shift $\sigma=0$ for the seed. 
Eigenpairs are obtained with the Rayleigh-Ritz projection method.

\begin{table}[tbp]
\begin{center}
\caption{The efficiency and the accuracy of the SS method with $C$ and $\mathbb{L}^{\pm}$ for calculating 6 low-lying eigenvalues and corresponding eigenvectors.}
\vspace*{3pt}
\begin{tabular}{c|c|ccc}
\noalign{\hrule height 1.0pt}
path & $C$ & \multicolumn{3}{c}{$\mathbb{L}^{\pm}$} \\ \hline
$\beta$ & - & 0.2 & 0.6 & 1.0 \\ \hline
\# matvec & 4352 & 4268 & 3560 & 2886 \\
time [sec] & 399.1 & 424.8 & 355.5 & 293.2 \\
res$_1$ & 1.8$\times 10^{-11}$ & 5.2$\times 10^{-10}$ & 3.1$\times 10^{-8}$ & 1.5$\times 10^{-6}$ \\
res$_2$ & 2.8$\times 10^{-10}$ & 1.8$\times 10^{-12}$ & 1.2$\times 10^{-9}$ & 3.0$\times 10^{-7}$ \\
res$_3$ & 4.0$\times 10^{-10}$ & 1.9$\times 10^{-12}$ & 1.6$\times 10^{-9}$ & 3.3$\times 10^{-7}$ \\
res$_4$ & 2.9$\times 10^{-10}$ & 3.0$\times 10^{-12}$ & 7.1$\times 10^{-10}$ & 1.1$\times 10^{-7}$ \\
res$_5$ & 5.6$\times 10^{-11}$ & 2.5$\times 10^{-12}$ & 4.0$\times 10^{-10}$ & 6.7$\times 10^{-8}$ \\
res$_6$ & 7.8$\times 10^{-12}$ & 6.6$\times 10^{-10}$ & 1.6$\times 10^{-8}$ & 5.8$\times 10^{-7}$ \\
\noalign{\hrule height 1.0pt}
\end{tabular}
\label{result1}
\end{center}
\end{table}

We show the efficiency and the accuracy of the SS method with $C$ and $\mathbb{L}^{\pm}$ in Table~\ref{result1},
where $\mathrm{res}_l \equiv ||A\textrm{\boldmath $x$}_l-\lambda_l\textrm{\boldmath $x$}_l||_2$, $||\textrm{\boldmath $x$}_l||_2=1$
is the residual for the $l$-th smallest eigenvalue $\lambda_l$.
In the case of $\mathbb{L}^{\pm}$, we obtain the lower accuracy of eigenpairs 
toward the edges of the integration interval as expected in Sec.~3.
We also find that accuracy of eigenpairs is increased
at the cost of 
the number of matrix-vector multiplications as $\beta$ becomes smaller.
This allows us to choose an optimal value of $\beta$ to minimize
the computational cost for the required precision of eigenpairs.  
We observe similar efficiency between the $C$ and 
$\mathbb{L}^{\pm}$ cases. An intriguing finding is that 
the elapsed time for $\mathbb{L}^{\pm}$ with $\beta=0.2$ is larger than that of $C$ even if the number of matrix-vector multiplications
of the former is less than that of the latter.
The reason is that the vector operations in the shifted CG method require
nonnegligible computational cost compared to the matrix-vector multiplication
which contains only 51 nonzero components in each row in our case.

We also compare the efficiency between PARPACK and 
the SS method with $C$ and $\mathbb{L}^{\pm}$ 
by calculating 20 low-lying eigenvalues and 
corresponding eigenvectors. 
The parameters of the SS method are chosen to satisfy the tolerance
$\mathrm{res}_l \leq 10^{-9}$ and the stopping criterion for PARPACK $tol = 10^{-10}$.
For the SS method with $C$ we employ 4 circles with $N=32$, $M=24$ 
choosing $\gamma=-0.02485,-0.00964,0.01181,0.02575$
and $\rho=0.00435,0.00960,0.00860,0.00431$, each of which contains
5 eigenvalues.
The SS method with $\mathbb{L}^{\pm}$ uses 6 pair of lines with $N=32$, $M=24$, $N^{\prime}=16$, $\rho=0.02121$ and $\beta=0.2$.
The other setup for the SS method is the same as the previous experiment.
For PARPACK the number of the Arnoldi vectors is chosen  
to be four times the number of eigenvalues, i.e. 80, in the regular mode. 

\begin{table}[tbp]
\begin{center}
\caption{The efficiency and the accuracy of the SS method with $C$ and $\mathbb{L}^{\pm}$ and PARPACK for calculating 20 low-lying eigenvalues and corresponding
eigenvectors.}
\vspace*{3pt}
\begin{tabular}{c|ccc}
\noalign{\hrule height 1.0pt}
 & SS ($C$) & SS ($\mathbb{L}^{\pm}$) & PARPACK \\
\noalign{\hrule height 1.0pt}
\# matvec & 7542 & 6680 & 7384 \\
Total \# quad. points & 128 & 102 & - \\
Total time [sec] & 1159.4 & 1020.7 & 1277.4 \\
Time for matvec [sec] & 421.2 & 365.5 & 406.1 \\
$\mathrm{res}_{\mathrm{max}}$ & 7.7$\times 10^{-10}$ & 2.6$\times 10^{-10}$ & 2.8$\times 10^{-12}$ \\
$\mathrm{res}_{\mathrm{min}}$ & 1.4$\times 10^{-12}$ & 7.7$\times 10^{-13}$ & 3.6$\times 10^{-15}$ \\
\noalign{\hrule height 1.0pt}
\end{tabular}
\label{result2}
\end{center}
\end{table}

Table \ref{result2} shows the efficiency and the accuracy of three methods.
$\mathrm{res}_{\mathrm{max}}$ and $\mathrm{res}_{\mathrm{min}}$ are the maximum and minimum value of $\mathrm{res}_l$, respectively.
There are two important points.
One is that the SS method shows similar or better
efficiency and accuracy 
in comparison with PARPACK thanks to the shifted CG method 
which reduces the number of matrix-vector multiplications by about 1/100. 
Another is that the SS method with $\mathbb{L}^{\pm}$ needs less numbers of
matrix-vector multiplications and quardrature points compared to the $C$
case. This is because the $\mathbb{L}^{\pm}$ case requires less number of iterations for the shifted CG method and allows us to reuse
the quadrature points.

\section{Conclusions}
We introduce a shifted Krylov subspace method to reduce the computational cost for the SS method.
Moreover we propose a new integration path along straight lines which decreases both the number of iterations for a Krylov subspace solver and
the number of quadrature points.

We calculate some low-lying eigenvalues and corresponding eigenvectors of a Hermitian fermion matrix with the SS method and PARPACK.
We show that the SS method becomes efficient comparable with PARPACK thanks to the shifted CG method and
our new integration path is more efficient than the conventional one on a circle.

Investigating more efficient integration paths and quadrature rules to reduce computational cost for the SS method is our future plan.

\section*{Acknowledgments}
Numerical calculations for the present work have been carried out on the T2K-TSUKUBA computer under the ``Interdisciplinary Computational Science
Program" of Center for Computational Sciences, University of Tsukuba. This work is supported in part by Grants-in-Aid for Scientific Research from
the Ministry of Education, Culture, Sports, Science and Technology (Nos. 18540250, 20105002, 21105502 and 21246018).

\end{document}